\documentclass[pra,aps,twocolumn,showpacs]{revtex4}
\usepackage{graphicx}
\usepackage{dcolumn}

\begin{document}
\def\be{\begin{equation}}
\def\ee{\end{equation}}
\def\bearr{\begin{eqnarray}}
\def\eearr{\end{eqnarray}}
\def\la{\langle}
\def\ra{\rangle}
\def\l{\left}
\def\r{\right}

\title{Dynamic structure factor of Fermi superfluid in the BEC-BCS crossover}

\author{Tarun Kanti Ghosh}
\affiliation
{Institute for Theoretical Physics, Heinrich-Heine Duesseldorf University,
40225, Duesseldorf, Germany}

\date{\today} 

\begin{abstract}
We consider cigar shaped Fermi superfluid in the BEC-BCS crossover.
Using polytropic form of equation of state, we derive low energy multibranch
bosonic excitations and the corresponding density fluctuations in three different 
regimes along the crossover, namely 
weak-coupling BCS, unitarity and molecular BEC regimes.
Bragg spectroscopy can be used to probe the multibranch nature of the low
energy bosonic excitations by measuring dynamic structure factor.
Therefore, we calculate dynamic structure factor in those three different regimes.
In the Bragg spectroscopy, an actual observable is momentum imparted to 
the superfluid due to the Bragg potential. We also present results of the momentum
imparted to the superfluid due to the Bragg pulses.  
\end{abstract}


\maketitle

\section{Introduction}
The crossover from Bose-Einstein condensation (BEC) to Bardeen-Cooper-Schrieffer (BCS)
state has drawn renewed interest in past few years due to rapid experimental progress 
in cold atomic two-component Fermi gases. Several experimental groups 
\cite{greiner,jochim, zw,hara,regal,barten,bourdel,chin} 
achieved the BEC-BCS crossover in the two-component atomic Fermi gases due to the 
magnetized Feshbach resonance mechanism \cite{houb,stwa,ties}.
The fermionic system becomes molecular BEC for strong repulsive interaction and 
transform into the BCS states when the interaction is
attractive. Near the resonance, the zero energy $s$-wave scattering length $a$
exceeds the interparticle spacing and the interparticle interactions
are unitarity limited and universal.

As in the case of bosonic clouds the frequencies of collective
modes of Fermi gases can be measured to high accuracy. The collective
oscillation frequencies of a trapped gas can provide crucial information 
on the equation of state of the system.
The experimental results on the collective frequencies of the lowest axial
and radial breathing modes on ultra cold gases of ${}^6$Li across the
Feshbach resonance have also become available \cite{freq1,freq2}. 
Since the weak-coupling BCS and unitarity limits are characterized by 
the same collective oscillation frequency, it is an interesting to find 
out another observable which makes a clear identification of these two regimes 
and to better characterize two kinds of superfluid.

All the experiments of cold atomic Fermi gases are done in a cigar shaped 
geometry in which the atomic density is inhomogeneous in the radial 
plane and quasi-homogeneous along the symmetry axis. The axial excitations 
of a cigar shaped Fermi superfluid can be divided into two regimes: 
short wavelength excitations whose wavelength is much smaller than 
the axial size and long wavelength excitations whose wavelength is 
equal or larger than the axial size of the system. 
In the later case, the axial excitations are discrete and the lowest
axial breathing mode frequency has been measured \cite{freq2}. 
In the former case, the short wavelength 
axial phonons with different number of radial modes of a cigar-shaped 
Fermi superfluid give rise to the multibranch spectrum \cite{zaremba}.
These are similar to the electromagnetic wave propagation in a wave guide.
In the BCS side of the resonance, the low energy bosonic excitations, apart 
from the gaped fermionic excitations, of the Fermi superfluid are called 
multibranch Bogoliubov-Anderson (BA) modes.
In the usual electronic superconductors, the BA phonon mode is absent due to
the long-range Coulomb interaction. 

In this work we find that the low energy multibranch modes in the BCS limit are
different from that of in the unitarity limit,
although discrete radial and axial mode frequencies are same due to the same exponent 
in the equation of state in the unitarity and BCS regimes.
Therefore, these multibranch modes can be used to identify and characterize the 
weak-coupling BCS and unitarity states. 

It is an interesting to study how one can probe such bosonic modes in
the current available experimental setup. In fact, these multibranch low-energy
bosonic modes could be observed by measuring dynamic structure factor (DSF) in a 
Bragg scattering experiment.
Bragg spectroscopy of a trapped atomic system has proven to be an important
tool for probing many bulk properties such as dynamic structure factor \cite{phonon}, 
verification of multibranch Bogoliubov excitation spectrum in the usual
atomic BEC \cite{mbs2}, correlation functions and momentum distributions of a phase 
fluctuating Bose gases \cite{ric,gerbier}.
There has been a suggestion of Bragg scattering experiment to study the
fermionic excitations at zero temperature \cite{deb} as well as at temperature 
close to the critical temperature \cite{baym}. 

In this work, we also calculate DSF in the various regimes of the crossover, namely 
the weak-coupling BCS, unitarity and molecular BEC regimes. In actual Bragg scattering 
experiments, the measured response function is momentum transferred
$ P_z(t) $ to the superfluid by the Bragg pulses. The momentum 
transferred is directly related to the dynamic structure factor when time duration of 
the Bragg pulses is long enough. Therefore, we also study momentum transferred to the 
superfluid by the Bragg pulses. 
In order to probe the multibranch nature of the modes, we also estimate required 
values of wave vector and time duration of the two-photon Bragg pulses in the Bragg
scattering experiments. 

Recently, there is a measurement of sound velocity along the crossover \cite{sound}.
The measured sound velocity do not match with the theoretical prediction in
the molecular BEC regime, however it matches very well in other regimes of the crossover
\cite{tkgsound}.
One can also estimate the sound velocity in the three different regimes along the
crossover by measuring slope of the phonon mode. 
It may resolve the puzzle of mismatch of the measured sound velocity with that of 
the theoretical prediction in the molecular BEC state.

This paper is organized as follows. In section II, we provide quantized hydrodynamic 
description of the Fermi superfluid along the crossover. In section III, we calculate
the dynamic structure factor along the crossover. In section IV, we discuss the 
possible Bragg scattering experiment in this system and present results of the momentum
transferred to the system due to the Bragg pulses.
We also discuss a summary and conclusions in section V.

\section{quantized hydrodynamic theory of Fermi superfluid}
We use hydrodynamic model with the Weizsacker quantum pressure term
to describe low-energy dynamics of a two-component Fermi superfluid at 
zero temperature. This system can be well described by a time-dependent 
non-linear Schrodinger equation as follows \cite{kim}: 
\be \label{sch}
i\hbar \frac{\partial \hat \psi ({\bf r},t)}{\partial t} = 
[- \frac{\hbar^2}{2M} \nabla^2  + V_{\rm ext}(r,z) + \bar \mu(n)] 
\hat \psi ({\bf r},t),
\ee
where the non-linear term 
$ \bar \mu(n) = \frac{\partial}{\partial n} [n \epsilon (n)] $ is the 
chemical potential in a uniform system and $ \epsilon (n) $ is the 
ground state energy per particle. Here, $ M $ is the mass of a Fermi 
atom and $ V_{\rm ext}(r,z) = (M/2)(\omega_{r}^2 r^2 +
\omega_z^2 z^2) $ is an external harmonic trap potential with 
$ \omega_{r} >> \omega_z $. We have taken $\omega_z/\omega_r = 0.1 $ and
total number of atoms $ N=2.0 \times 10^{6}$ in all numerical calculations.
On the basis of quantum Monte Carlo data of Astrakharchik {\em et al.}
\cite{astra}, 
Manini and Salasnich \cite{manini} proposed a very useful analytical 
fitting expression for $ \epsilon (n) $ (see Eq. (6) and Table 1 of Ref. 
\cite{manini}). 
The Hamiltonian corresponds to Eq. (\ref{sch}) can be written as
$$
H = \int d {\bf r} \hat \psi^*({\bf r},t)[- \frac{\hbar^2}{2M} \nabla^2 
+ V_{\rm ext}(r,z) +  \bar \mu(n)] \hat \psi ({\bf r},t).
$$
Using phase ($\theta $)-density ($n$) representation of the order parameter of 
the composite bosons: 
$ \hat \psi ({\bf r},t) = \sqrt{\hat n ({\bf r},t)} e^{i\hat \theta ({\bf r},t)} $, 
the above Hamiltonian becomes,

\bearr
\hat H & =  & \int d {\bf r}[\sqrt{\hat n}(- \frac{\hbar^2}{2M} \nabla^2 \sqrt{\hat n})
+ \frac{1}{2} M {\hat {\bf v}}^2 n + V_{\rm ext}(r,z) \hat n \nonumber \\ 
& + &  \bar \mu(n) \hat n].
\eearr

Linearizing the density and phase around their equilibrium values:
$ \hat n({\bf r},t) = n_0({\bf r}) + \delta \hat n({\bf r},t) $ and 
$ {\hat {\bf v}}({\bf r},t) = \delta {\hat {\bf v}}({\bf r},t) $,
where $ {\hat {\bf v}} = (\hbar/M) \nabla \hat \theta $ is the superfluid 
velocity.
Keeping upto quadratic fluctuations, the above Hamiltonian reads

\be
H = H_0 + \int d{\bf r}  [\frac{M \delta {\hat {\bf v}}^2}{2} 
  + \frac{\partial \bar \mu}{\partial n}|_{n_0} \delta \hat n^2],
\ee
where $ H_0 $ is the ground state energy. By using time-dependent 
Heisenberg equations of motion for the density and velocity fluctuations,
one can get continuity and Euler's equations which are given by

\be \label{den}
\delta \dot {\hat n} = - \nabla \cdot[n_0(r) \delta \hat {\bf v}],
\ee
and 
\be \label{vel}
M \delta \dot {\hat {\bf v}} = - \nabla [ \frac{\partial \bar \mu}
{\partial n}|_{n_0} \hat \delta n].
\ee

Since the Hamiltonian is quadratic in terms of the fluctuation operators, 
it can be diagonalized
by using the following standard canonical transformations:
\be \label{dencan}
\delta \hat n (r,z,t) = \sum_{j,k} [ A_{j,k} \psi_{j,k}(r) 
\hat \alpha_{j,k} 
e^{i(kz- \omega_j(k) t)} + h. c.], 
\ee
and
\be
\delta \hat \theta (r,z,t) = \sum_{j,k} [ B_{j,k} \psi_{j,k}(r) \hat \alpha_{j,k} 
e^{i(kz- \omega_j(k) t)} + h. c.].
\ee
Here, $j$ is a set of two quantum numbers: radial quantum number,
$n_r$ and the angular quantum number, $m$. Also, $ k$ is the axial wave vector.
The density and phase fluctuations satisfy the following equal-time commutator
relation: $  [\delta \hat n(r), \delta \hat \theta(r^{\prime})] 
= i \delta (r - r^{\prime}) $.
One can easily show that
\be
A_{j,k} = i \sqrt{\frac{\hbar \omega_j(k)}{2 \frac{\partial \bar \mu}
{\partial n}|_{n_0}}},
B_{j,k} = \sqrt{\frac{2 \frac{\partial \bar \mu}{\partial n}|_{n_0}}
{\hbar \omega_j(k)}}
\ee
We are assuming that $ \psi_{j,k}(r) $ satisfies the orthonormal
conditions:
$ \int d {\bf r} \psi_{j,k}^*(r) \psi_{j^{\prime},k}(r) = 
\delta_{jj^{\prime}} $
and $ \sum_j \psi_{j,k}(r) \psi_{j,k}^*({ r}^{\prime}) = 
\delta (r -  r^{\prime}) $.

We assume power-law form of the equation of state as
$ \bar \mu(n) = C n^{\gamma} $. Here, $C$ depends on interaction strength and 
the effective polytropic index $ \gamma $ is a function of a dimensionless parameter 
$ y = 1/k_F a$, where $k_F$ is the Fermi wave vector and $ a $ is the scattering 
length between Fermi atoms of different components. The 
weak-coupling BCS ($y<<-1$) 
and the unitarity ($y=0$) states are described by the same exponent $ \gamma = 2/3 $ with 
different values of $C$. For the weak-coupling BCS regime, 
$ C_{\rm bcs} \simeq  (3\pi^2)^{2/3}(3\hbar^2/10M) $ and
for the unitarity regime, $C_{\rm uni} = 0.44 C_{\rm bcs} $.  
The molecular 
BEC state ($y>>1$) is described by $\gamma = 1$ and 
$ C_{\rm bec} = 4 \pi \hbar^2 a_m/M $, where 
$ a_m = 0.6 a$ is the molecular scattering length \cite{gvs}.
In our calculation we have
assumed $ a_m = 1.0 \times 10^{-8} m $.
The power-law form of the equation of state is being used successfully to study the 
Fermi superfluid along the crossover \cite{manini,poly1,poly2,bulgac,astra1}. 
At equilibrium, the density profile takes the form
$ n_0(r) = (\mu/C)^{1/\gamma}( 1- \tilde r^2)^{1/\gamma} $,
where $ \tilde r = r/R_{0} $,
$ R_{0} = \sqrt{2 \mu/M \omega_{r}^2} $ and $ \mu $ is the chemical potential
in the non-uniform system, which can be obtained from the normalization condition.

Taking first-order time-derivative of Eq. (\ref{den}) and using Eq. (\ref{vel}),
the second-order equation of motion for the density fluctuation is
given by
\be \label{den0}
\frac{\partial^2 \delta n}{\partial t^2} =
\nabla \cdot [n_0(r) \nabla \frac{\partial \bar \mu(n) }{\partial n}|_{n =n_0}
\delta n ].
\ee

Using the polytropic form of the equation of state and Eq. (\ref{dencan}),
then Eq. (\ref{den0}) reduces to the following equation: 

\bearr \label{den1}
- \tilde \omega_{j}^2(k) \psi_{j,k}(r) & = & [\frac{\gamma}{2} \nabla_{\tilde r}
\cdot [(1- \tilde r^2)^{1/\gamma}
\nabla_{\tilde r}(1- \tilde r^2)^{1-1/\gamma}] \nonumber \\
& - &  \frac{\gamma}{2} \tilde k^2 (1-\tilde r^2)] \psi_{j,k}(r),
\eearr
where $ \tilde \omega = \omega/\omega_r $ and $ \tilde k = k R_{0} $.

For $ k = 0 $, it reduces to a two-dimensional eigenvalue problem and the solutions
of it can be obtained analytically. The energy spectrum is
given by
\be
\tilde \omega_{j}^2 = |m| + 2 n_r [\gamma(n_r + |m|) + 1].
\ee   
The corresponding orthogonal eigenfunction is given by
\be
\psi_{j} \propto (1-\tilde r^2)^{1/\gamma -1} \tilde r^{|m|}
P_{n_r}^{(1/\gamma -1, |m|)} (2\tilde r^2 -1) e^{im\phi},
\ee
where $ P_{n}^{(a,b)}(x) $ is a Jacobi polynomial of order $n$ and $\phi $ is
the polar angle.

The solution of Eq. (\ref{den1}) 
can be obtained for arbitrary value of $k$ by
numerical diagonalization.
For $ k \neq 0 $, we expand the density fluctuation as
\be
\psi_{j,k}(r,\phi) = \sum_{j} b_{j} \psi_{j}(r,\phi).
\ee  

Substituting the above expansion into Eq. (\ref{den1}), we obtain,
\bearr \label{density2}
0 & = & [\tilde \omega_{j}(k)^2 - [|m| + 2 n_r ( \gamma (n_r +|m|) +1)]
\nonumber \\ & - &
\frac{\gamma}{2} \tilde k^2] b_{j} +
\frac{\gamma}{2} \tilde k^2 \sum_{j^{\prime}} M_{j j^{\prime}}
b_{j^{\prime}}.
\eearr
Here, the matrix element $ M_{j j^{\prime}} $ is given by
\be \label{matrix}
M_{j j^{\prime}} = \int d^2 \tilde r \psi_{j}
\tilde r^2 \psi_{j^{\prime}}.
\ee
The above eigenvalue problem (Eq. (\ref{density2})) is block diagonal with 
no overlap between the subspaces of different angular momentum, so that the 
solutions to Eq.(\ref{density2}) can be obtained separately in each angular 
momentum subspace. We can obtain all low energy multibranch spectrum on the 
both sides of the Feshbach resonance including the unitarity 
limit from Eq. (\ref{density2}). Equations (\ref{density2}) and (\ref{matrix}) 
show that the spectrum depends on the average over the radial coordinate and the 
coupling between the axial mode and transverse modes within a given angular 
momentum symmetry. Particularly, the coupling is important for large values of 
$k $. 
We are interested to study $m=0$ states since these states are
excited in the Bragg scattering experiments due to axial symmetry of the system. 
We show low energy multibranch modes of $m=0$ states in three different regimes 
in Fig. 1. The
top three panels of Fig. 1 show the multibranch spectrum in the three different
regimes. To compare these spectrum, we have plotted all those spectrum of 
different regimes in a single frame, which is shown in the bottom panel of Fig. 1.
The lowest branch corresponds to the Bogoliubov axial mode with no radial nodes.
This mode has the usual form $ \omega = c_s k $ at low momenta, where 
$c_s $ is the sound velocity. In the limit of small $k$, the other branches have 
free-particle dispersion due to the gaped nature of these modes.

The discrete radial and axial modes are same in the unitarity and BCS regimes 
since the exponent $ \gamma $ are the same for both the regimes. However, the 
multibranch modes are different in the unitarity and BCS regimes in spite of 
the same exponent in the equation of state. This is due to the different radial 
sizes in those two regimes for a given number of atoms and the trap potential. 
Therefore, these low energy axial propagation of discrete radial modes can be 
used to characterize different regimes of the superfluid.   
The density fluctuations for a fixed value of the axial momentum are plotted in 
Fig. 2. The density fluctuations corresponds to the multibranch modes are also 
different in the different regimes. In Fig. 2, the magnitude of the density fluctuations
are given in an arbitrary unit since these are the linear fluctuations. 
\begin{figure}[ht]
\includegraphics[width=8.0cm]{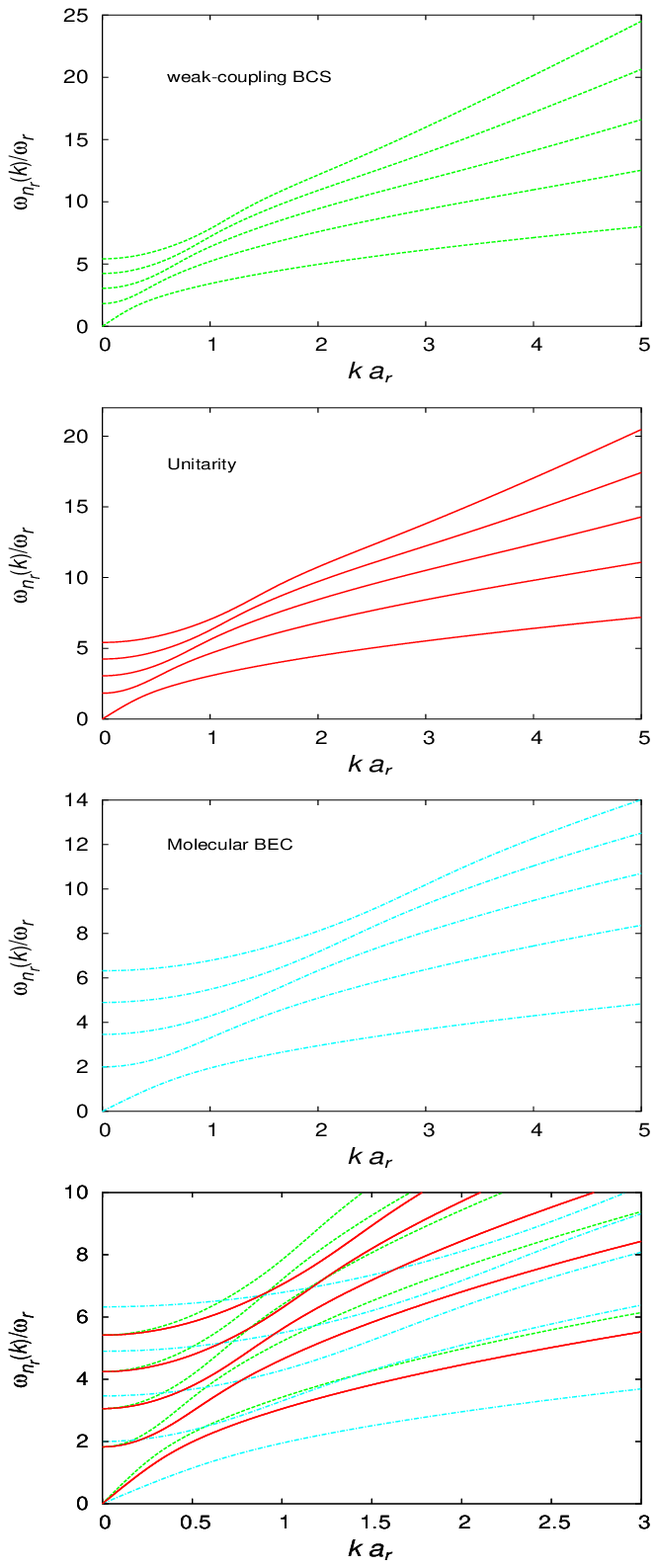}
\caption{(Color online) Plots of the low energy multibranch modes in the three different 
regimes.}
\end{figure}

\section{dynamic structure factor}
The dynamic structure factor is the Fourier transformation of density-density
correlation functions and it is given as

$$
S(k,\omega) = \int d {\bf r} d {\bf r}^{\prime} dt 
<\delta \hat n^{\dag}({\bf r},t) \delta \hat n ({\bf r}^{\prime},0)> 
e^{i{\bf k} \cdot ({\bf r} - {\bf r}^{\prime})} e^{i\omega t} 
$$

This can be written as
\be \label{delta} 
S(k,\omega) =  \sum_j S_{j}(k) \delta (\omega - \omega_j(k)),
\ee
where the weight factor $ S_{j}(k) $ is given by
\be
S_{j}(k) = \frac{\hbar \omega_j(k)}{2 \frac{\partial \bar \mu}{\partial n}|_{n_0} } 
|\psi_{j}(k)|^2.
\ee
Here, 
$ \psi_j(k) = \int d {\bf r} e^{-i{\bf k} \cdot {\bf r}} \psi_{j,k}({r})$ is the 
Fourier transform of the eigenfunctions $ \psi_{j,k}({r})$. 
The weight factors are plotted in Fig. 3.
The weight factors $ S_{j}(k) $ 
determine how many modes are excited for a given value of $k$.
For example, when $ ka_{r} = 1.0 $,  $ n_r $ = \{0, 1\}, $ n_r $ = \{0, 1, 2 \} and
$ n_r $ = \{0, 1, 2, 3\} modes are excited in the molecular BEC, unitarity and
weak-coupling BCS regimes, respectively.
The harmonic oscillator length is defined as $ a_r = \sqrt{\hbar/M \omega_r} $.
To excite many other low-energy modes, the wave vector in the Bragg potential must be large.
From Fig. 3, it is clear that $ S_0 (k) \sim k $ and $ S_{n_r>0} (k) $ is almost constant 
in the limit of small $k$. Therefore, $ \omega_0 (k) \sim k $ and
$ \omega_{n_r>0} (k) \sim k^2 $ when $k $ is very small.
Therefore, our analysis also satisfies the Feynman-like relation
$\omega_{n_r}(k) \sim k^2/S_{n_r}(k) $.
The dynamic structure factors of the three different regimes for $ k a_{r} = 0.5 $ are 
plotted in Fig. 4. The delta function in Eq. (\ref{delta}) is replaced by the Lorentzian form to 
plot Fig. 4.
\begin{figure}[ht]
\includegraphics[width=8.0cm]{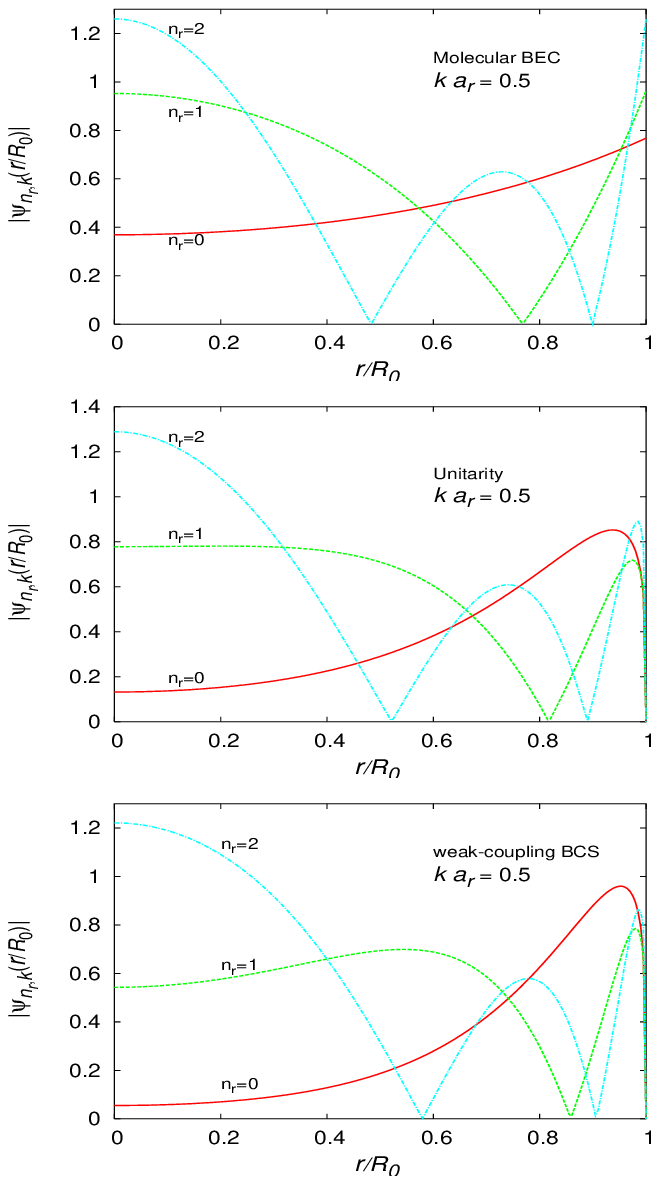}   
\caption{(Color online) Plots of the low-energy density fluctuations in the three different 
regimes
for $ k a_{r} = 0.5 $.}
\end{figure}

\begin{figure}[ht]
\includegraphics[width=8.0cm]{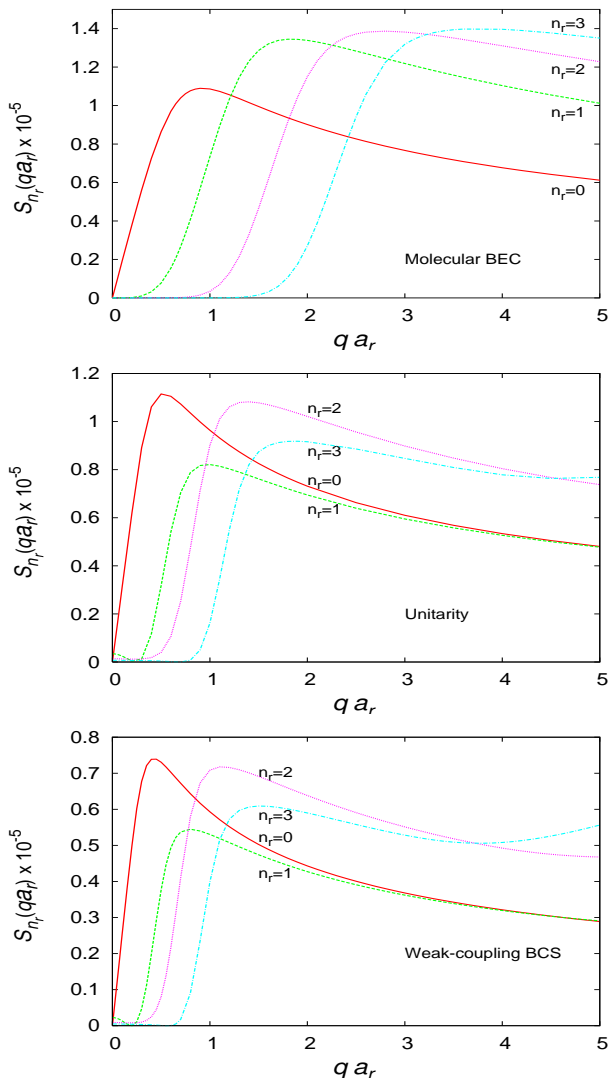}
\caption{(Color online) Plots of the weight factors in the three different regimes.}
\end{figure}

\begin{figure}[ht]
\includegraphics[width=8.0cm]{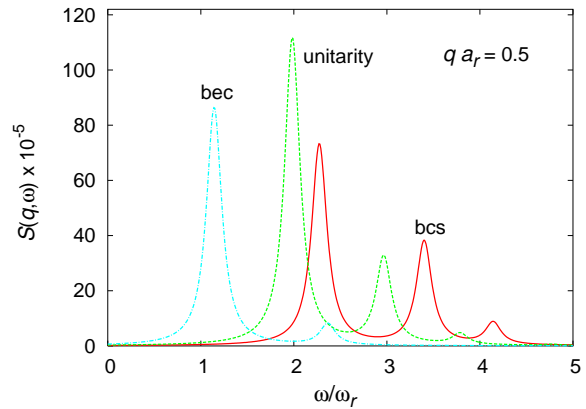}
\caption{(Color online) Plots of the dynamic structure factors of the three different
regimes for $ k a_{r} = 0.5$.}
\end{figure}

\section{Bragg scattering experiment}
The behavior of these multiple peaks in the dynamic structure factor can 
be resolved in a two-photon Bragg spectroscopy, as shown by Steinhauer 
{\em et al.} \cite{davidson} for usual BEC. In the two-photon Bragg spectroscopy, the 
dynamic structure factor can not be measured directly. Actually, the observable 
in the Bragg scattering experiments is the momentum transferred to the 
superfluid, which is related to the dynamic structure factor and reflects
the behavior of the quasiparticle energy spectrum.
The populations in the quasiparticle states can be controlled by using
the two-photon Bragg pulse. When the superfluid is irradiated by an external 
moving optical potential $ V_{\rm op} = V_B(t) \cos(qz-\omega t)$, the excited states 
are populated by the quasiparticle with energy $ \hbar \omega $ and the momentum 
$\hbar q$, depending on the value of $q$ and $\omega$ of the optical potential 
$ V_{\rm op} $. Here, $ V_B $ is the intensity of the Bragg pulse.
Suppose the system is subjected to a time-dependent Bragg pulse which is switched 
on at time $t>0$ and $q$ is also along the $z$-direction.
We calculate, similar to the calculation of Refs. \cite{tkg,blak}, the momentum 
transfer to the superfluid from the moving optical potential 
and it is given by
\bearr
P_z(t) & = & \sum_{j,k} \hbar k < \hat \alpha_{j,k}^{\dag} (t) \hat \alpha_{j,k} (t) >
 =   \l (\frac{V_B(t)}{2 \hbar} \r )^2  \nonumber \\ 
& \times & \sum_j \hbar q S_j (\tilde q)    
\times  [F_j (\omega_{-}t) - F_j (\omega_{+}t)],
\eearr
where $ \hat \alpha_{j,k} (t) $ is the time-evolution of the quasiparticle operator 
of energy $\hbar \omega_j(k) $ and  
\be
F_j (\omega_{\pm}t) =
\l (\frac{\sin[(\omega_j(q) \pm \omega)t/2]}{(\omega_j(q) \pm \omega)/2} \r )^2. 
\ee
For positive $ \omega $ and a given $ \tilde q $ such that $ S_j(\tilde q) $ is maximum,
the momentum transferred $ P_z(t)$ is resonant at the frequencies $ \omega = \omega_j(q) $.
The width of the each peak goes like $ 2 \pi/t $. 
For large $t$ and $ \omega_z << \omega_{r} $, one can show that 
$P_z(t) \sim  S(k,\omega)$ \cite{tozo}.

In Fig. 5, we plot the net momentum transfer $P_z(t)$ vs the Bragg frequency
$ \omega $ for three different choices of the time duration of the Bragg pulses.
Figure 5 shows that the shape of the $P_z(t)$ strongly depends on the time duration  
of the Bragg pulses $ t_B$.
When $ t_B = 0.5 T_r $, $P_z(t)$ is a smooth curve with a single peak.
Here, we define $ T_{r} = 2 \pi/\omega_{r}$ is radial trapping period.
When $t_B = 1.0 T_r $, there is a little evidence of few small peaks start developing in 
$P_z(t)$.
When $ t_B = 2.0 T_r $, the multiple peaks in $P_z(t)$ appears prominently.
Therefore, the duration of the Bragg pulses $t_B $ should be greater than the radial
trapping period $T_r$ in order to resolve different peaks in the DSF.
Figure 5 also shows that the number of quasiparticle modes in unitarity regime is
much higher than the other regimes of the crossover. This is due to the fact 
that $ P_z(t) $ is proportional to the number of quasiparticle modes for a given $k$.
\begin{figure}[ht]
\includegraphics[width=8.0cm]{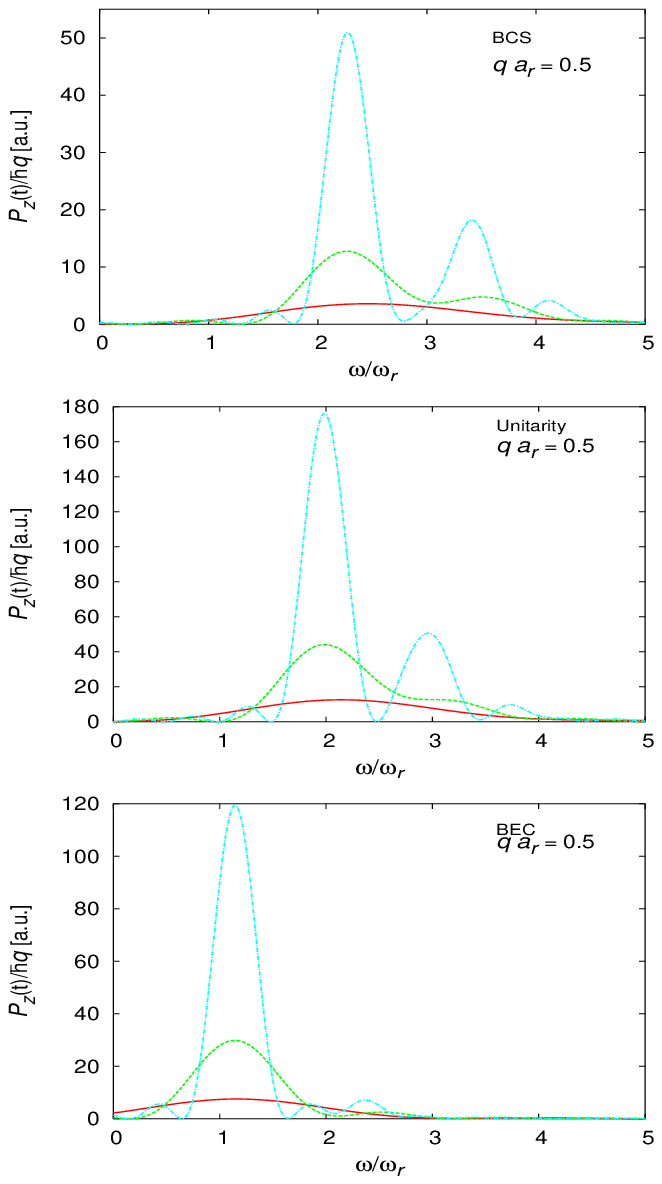}
\caption{(Color online) Plots of the momentum transferred 
in the three different regimes for $k a_{r} = 0.5$ and for different
time duration of the Bragg pulses: $t=0.5 T_r$ (solid), 
$t=1.0 T_r$ (dashed) and $t=2.0 T_r$ (dot-dashed).}
\end{figure}

\section{summary and conclusions}
We have presented the quantized hydrodynamic theory of cigar shaped Fermi 
superfluid along the BEC-BCS crossover by using the power law form of the 
equation of state. We have calculated multibranch low energy bosonic modes and 
the corresponding density fluctuations in three different regimes along the BEC-BCS 
crossover, namely the weak-coupling BCS, unitarity and molecular BEC states. Then 
we have presented results of the dynamic structure factor calculation. We have also 
calculated the momentum transferred to the superfluid by the Bragg pulses and shown 
that the multibranch nature would be observed when the time duration of the Bragg pulses
is greater than the radial trapping period. 

We have found that the axial propagation of discrete radial modes in the weak-coupling 
BCS and unitarity regimes are different, although the axial and radial modes 
are same in both cases. One can identify these two regimes by probing these multibranch 
modes with the help of the Bragg scattering experiment. 
We have also seen that the number of quasi particle modes in the unitarity regime is quite 
large compared to other two regimes for a fixed value of the Bragg momentum and the number 
of atoms. Therefore, the response function in the unitarity regime will be more prominent 
than the other two regimes in the crossover. 
Moreover, one can estimate the sound velocities in different regimes along the crossover by 
measuring the slope of the phonon modes.

\begin{acknowledgments}
This work was supported by the Alexander von Humboldt foundation, Germany.
\end{acknowledgments}


\begin{thebibliography}{2}
\bibitem{greiner}
M. Greiner, C. A. Regal, and D. S. Jin, Nature {\bf 426}, 537 (2003).

\bibitem{jochim}
S. Jochim, M. Bartenstein, A. Altmeyer, G. Hendl, S. Riedl,
C. Chin, J. H. Denschlag, and R. Grimm, Science {\bf 302}, 2101 (2003).

\bibitem{zw}
M. W. Zwierlein, C. A. Stan, C. H. Schunck, S. M. F. Raupach, S. Gupta,
Z. Hadzibabic, and W. Ketterle, Phys. Rev. Lett. {\bf 91}, 250401 (1993).

\bibitem{regal}
C. A. Regal, M. Greiner, and D. S. Jin, Phys. Rev. Lett. {\bf 92}, 040403 (2004).

\bibitem{barten}
M. Bartenstein, A. Altmeyer, S. Riedl, S. Jochim, C. Chin, J. H.
Denschlag, and R. Grimm, Phys. Rev. Lett. {\bf 92}, 120401 (2004).

\bibitem{bourdel}
T. Bourdel, L. Khaykovich, J. Cubizolles, J. Zhang, F. Chevy,
M. Teichmann, L. Tarruell, S. J. J. M. F. Kokkelmans, and
C. Salomon, Phys. Rev. Lett. {\bf 93}, 050401 (2004).

\bibitem{chin}  
C. Chin, M. Bartenstein, A. Altmeyer, S. Riedl, S. Jochim, J. H. Denschlag,
and R. Grimm, Science {\bf 305}, 1128 (2004).

\bibitem{hara}
K. M. O'Hara, S. L. Hemmer, M. E. Gehm, S. R. Grande,
and J. E. Thomas, Science {\bf 298}, 217 (2002).

\bibitem{houb}
M. Houbiers, H. T. C. Stoof, W. I. McAlexander, and R. G. Hulet,
Phys. Rev. A {\bf 57}, R1497 (1998).

\bibitem{stwa}
W. C. Stwalley, Phys. Rev. Lett. {\bf 37}, 1628 (1976).

\bibitem{ties}
E. Tiesinga, B. J. Verhaar, and H. T. C. Stoof, Phys. Rev. A {\bf 47},
4114 (1993).

\bibitem{freq1}
J. Kinast, S. L. Hemmer, M. E. Gehm, A. Turlapov, and J. E. Thomas,
Phys. Rev. Lett. {\bf 92}, 150402 (2004).

\bibitem{freq2}
M. Bartestein, A. Altmeyer, S. Riedl, S. Jochim, C. Chin, J. H. Denschlag, and
R. Grimm, Phys. Rev. Lett. {\bf 92}, 203201 (2004).

\bibitem{zaremba}
E. Zaremba, Phys. Rev. A {\bf 57}, 518 (1998).

\bibitem{phonon}
J. Stenger, S. Inouye, A. P. Chikkatur, D. M. Stamper-Kurn, D. E. Pritchard, and W. 
Ketterle,
Phys. Rev. Lett. {\bf 82}, 4569 (1999);
D. M. Stamper-Kurn, A. P. Chikkatur, A. Gorlitz, S. Inouye, S. Gupta,
D. E. Pritchard, and W. Ketterle, Phys. Rev. Lett. {\bf 83}, 2876 (1999).

\bibitem{mbs2}
J. Steinhauer, N. Katz, R. Orezi, N. Davidson,
C. Tozzo and F. Dalfovo, Phys. Rev. Lett. {\bf 90}, 060404 (2003).

\bibitem{ric}
S. Richard, F. Gerbier, J. H. Thywissen, M. Hugbart, P. Bouyer, and A. Aspect,
Phys. Rev. Lett. {\bf 91}, 010405 (2003).

\bibitem{gerbier}
F. Gerbier, J. H. Thywissen, S. Richard, M. Hugbart, P. Bouyer, and A. Aspect,
Phys. Rev. A {\bf 67}, 051602 (2003).

\bibitem{deb}
B. Deb, J. Phys. B {\bf 39}, 529 (2006).

\bibitem{baym}
G. M. Bruun and G. Baym, Phys. Rev. A {\bf 74}, 033623 (2006).

\bibitem{sound}
J. Joseph, B. Clancy, L. Luo, J. Kinast, A. Turlapov, and J. E. Thomas,
Phys. Rev. Lett. {\bf 98}, 170401 (2007).

\bibitem{tkgsound}
T. K. Ghosh and K. Machida, Phys. Rev. A {\bf 73}, 013613 (2006).

\bibitem{kim}
Y. E. Kim and A. L. Zubarev, Phys. Rev. A {\bf 70}, 033612 (2004).

\bibitem{astra}
G. E. Astrakharchik, J. Boronat, J. Casulleras, and S. Giorgini, Phys. Rev. Lett. {\bf 93}, 
200404 (2004).

\bibitem{manini}
N. Manini and L. Salasnich, Phys. Rev. A {\bf 71}, 033625 (2005).

\bibitem{gvs}
D. S. Petrov, C. Salomon, and G. V. Shlyapnikov, Phys. Rev. Lett. {\bf 93},
090404 (2004).

\bibitem{poly1}
H. Heiselberg, Phys. Rev. Lett. {\bf 93}, 040402 (2004).

\bibitem{poly2}
H. Hu, A. Minguzzi, X. J. Liu, and M. P. Tosi, Phys. Rev. Lett. {\bf 93}, 
190403 (2004).

\bibitem{bulgac}
A. Bulgac and G. F. Bertsch, Phys. Rev. Lett. {\bf 94}, 070401 (2005).

\bibitem{astra1}
G. E. Astrakharchik, R. Combescot, X. Leyronas, and S. Stringari,
Phys. Rev. Lett. {\bf 95}, 030404 (2005).

\bibitem{davidson}
J. Steinhauer, N. Katz, R. Ozeri, N. Davidson, C. Tozzo, and F. Dalfovo,
Phys. Rev. Lett. {\bf 90}, 060404 (2003).

\bibitem{tkg}
T. K. Ghosh, Int. J. Mod. Phys. B {\bf 20}, 5443 (2006).

\bibitem{blak}
P. B. Blakie, R. J. Ballagh, and C. W. Gardiner, Phys. Rev. A {\bf 65}, 
033602 (2002).

\bibitem{tozo}
C. Tozzo and F. Dalfovo, New J. Phys. {\bf 5}, 54 (2003).

\end{thebibliography}
\end{document}